\documentclass[12pt,preprint]{aastex}

\usepackage{amsmath}
\usepackage{amsbsy}
\usepackage{color}

\slugcomment{{\today}}

\shorttitle{Sympathetic Breakout CMEs from Pseudostreamers}
\shortauthors{Lynch and Edmondson}

\begin{document}


\title{Sympathetic Magnetic Breakout Coronal Mass Ejections from Pseudostreamers}


\author{B. J. Lynch\altaffilmark{1} and J. K. Edmondson\altaffilmark{2}}

%

%
\affil{\altaffilmark{1}Space Sciences Laboratory, University of California, Berkeley, CA 94720, USA; blynch@ssl.berkeley.edu}
\affil{\altaffilmark{2}Atmospheric, Oceanic and Space Sciences Department, University of Michigan, Ann Arbor, MI, 48109, USA; jkedmond@umich.edu}

\begin{abstract}

We present high resolution 2.5-dimensional MHD simulation results of magnetic breakout-initiated coronal mass ejections (CMEs) originating from a coronal pseudostreamer configuration.
The coronal null point in the magnetic topology of pseudostreamers means the initiation of consecutive sympathetic eruptions is a natural consequence of the system's evolution.
A generic source region energization process -- ideal footpoint shearing parallel to the pseudostreamer arcade polarity inversion lines -- is all that is necessary to store sufficient magnetic energy to power consecutive CME eruptions given that the pseudostreamer topology enables the breakout initiation mechanism. The second CME occurs because the eruptive flare reconnection of the first CME simultaneously acts as the overlying pre-eruption breakout reconnection for the sympathetic eruption.
We examine the details of the magnetic and kinetic energy evolution and the signatures of the overlying null point distortion, current sheet formation, and magnetic breakout reconnection giving rise to the runaway expansion that drives the flare reconnection below the erupting sheared field core.
The numerical simulation's spatial resolution and output cadence are sufficient to resolve the formation of magnetic islands during the reconnection process in both the breakout and eruptive flare current sheets. 
We quantify the flux transfer between the pseudostreamer arcades and show the eruptive flare reconnection processes flux $\sim$10 times faster than the pre-eruption breakout reconnection.
We show that the breakout reconnection jets cause bursty, intermittent upflows along the pseudostreamer stalk as well as downflows in the adjacent pseudostreamer arcade, both of which may be observable as pre-eruption signatures. 
Finally, we examine the flux rope CME trajectories and show that the breakout current sheet provides a path of least resistance as an imbalance in the surrounding magnetic energy density and results in a non-radial CME deflection early in the eruption.

\end{abstract}


%
%
\section{Introduction}

The weak polar fields during the declining phase of solar activity cycle 23 and the onset of cycle 24 \citep{Munoz2012} has meant a more highly warped coronal helmet streamer belt \citep{Luhmann2009} and consequently, the existence of more pseudostreamers \citep{Riley2012}. A pseudostreamer is characterized by the coronal arcade above an even number of polarity inversion lines (PILs) in 2-dimensional axisymmetric configurations \citep[e.g.,][]{Wang2007,Wang2012} or a single, connected PIL surrounding a magnetic region of the opposite polarity than the large-scale background field in 3-dimensions. Given the complexity of PIL distributions on the photosphere, pseudostreamer geometries in 3D can range from circular ``anemone" configurations in coronal holes \citep[e.g.,][and references therein]{Asai2008} to extended quasi-2D structures overlying adjacent high-latitude filament channels \citep[e.g.,][]{Panasenco2010,Torok2011,Titov2012}. 
The increase in high resolution, high cadence imaging data and overall longitudinal coverage from the STEREO and SDO missions means we are seeing with unprecedented detail the timing, location, evolution and connectivity of transient solar phenomena, including what has become known as ``sympathetic" eruptions \citep[e.g.,][]{Schrijver2011,Balasubramaniam2011,Yang2012,Shen2012}.

The combination of these two factors has contributed to community interest in pseudostreamers as CME source regions.
For example, two of the recent working groups at the 2012 NSF Solar and Heliospheric INterplanetary Environment (SHINE) Workshop were entitled ``Sympathetic and Homologous Eruptions in the Solar Corona" and ``CME Triggers: Flux Emergence, Topology, and Global Environment" and included significant discussion about the magnetic topology of CME source regions, CME initiation mechanisms, and their interrelation and interdependence.

While there has been some previous statistical studies of CME properties originating from pseudostreamer source regions \citep[][]{Liu2007}, recent modeling \citep{Torok2011,Zuccarello2012} has highlighted this topological aspect and started to investigate the consequences of the source environment on the eruptions. 
For example, \citet{Torok2011} have discussed the viability of the breakout mechanism for CME initiation from pseudostreamers, but 
given the similarity between the magnetic configuration of a pseudostreamer and the topological features necessary for magnetic breakout CME initiation \citep{Antiochos1999,DeVore2008,Lynch2008}, we have performed a high resolution 2.5D MHD simulation in order to demonstrate that a completely generic energization scenario can result in the eruption of sympathetic magnetic breakout CMEs. 
It has been an open question whether breakout-initiated CMEs under streamer configurations (i.e., breakout reconnection with overlying open fields) could produce a fast eruption. The simulation results discussed by \citet{Lynch2011} included a $\sim$1000~km~s$^{-1}$ breakout CME from under the helmet streamer belt in a realistic background solar wind. The results presented herein show that even the pseudostreamer's secondary, sympathetic CME has strong eruptive flare reconnection that accelerates the CME with virtually no overlying closed flux. Our simulation is consistent with 
the recent \citet{Karpen2012} results that highlight the role of the eruptive flare reconnection in providing the transition from a ``slow" to ``fast" eruption.

The paper is organized as follows. In Section~\ref{sec_init}, we present a brief description of the numerical model, the initial magnetic field configuration and plasma properties, and the form of the energizing shear flows. In Section~\ref{sec_res}, we present an overview of the sympathetic eruption scenario in the simulation results, examine the energy storage and release, and discuss the details of the role of the coronal current sheets and subsequent magnetic reconnection in the sympathetic eruption process. In Section~\ref{sec_disc}, we discuss the limitations of our simulation but also describe some of the observational consequences arising from the pseudostreamer source region, both in terms of pre-eruption signatures of the arcade/streamer stalk evolution and in the early dynamics of the flux rope eruption. In Section~\ref{sec_conc}, we conclude with a summary of our findings.


\section{Initial Conditions and Energizing Flows}
\label{sec_init}

The numerical simulations were run with the Adaptively Refined MHD Solver (ARMS) code, developed by C. Richard DeVore and collaborators at the U.S. Naval Research Laboratory. ARMS calculates solutions to the 3D nonlinear, time-dependent MHD equations that describe the evolution and transport of density, momentum, and energy throughout the plasma and the evolution of the magnetic field and electric currents \citep[see, e.g.][]{DeVore2008,Karpen2012}. The numerical scheme used is a finite volume, multi-dimensional flux-corrected transport algorithm \citep{DeVore1991}. 
The ARMS code is fully integrated with the adaptive mesh toolkit PARAMESH \citep{MacNeice2000} to handle dynamic, solution-adaptive grid refinement and support efficient multi-processor parallelization.

In our Cartesian axisymmetric geometry, the $z$-coordinate is the direction of translational symmetry. Therefore, in the MHD equations, we solve for the $z$-component of the vector quantities (velocity, magnetic field, current density) but there is no spatial variation of any of the physical quantities in the $z$ direction.  
The full computational domain is made up of 16 $\times$ 32 blocks in the $x$, $y$ directions with 8 cells per side, and 3 additional levels of static grid refinement for a effective maximum resolution of 1024$\times$2048. 
The full computation domain extends from $x \in \{-5.0,5.0\}$, $y \in \{1.0,21.0\}$ where the characteristic length scale is $\langle L_0 \rangle = 10^{9}$~cm. Since we are interested in the dynamics of the magnetic field and plasma of the pseudostreamer arcades, we will focus herein on the lower half of the computational domain ($y \le11.0$).
We use periodic boundary conditions at the $x = \pm 5.0$ walls, the lower $y$ boundary is line tied, and the upper $y$ boundary is open. We will argue in Section~4 that, while the periodicity has an obvious impact on the simulation results, the overall eruption scenario is both generic and robust, and would therefore proceed in a similar manner as presented here.

\begin{figure}
\center \includegraphics[width=24pc]{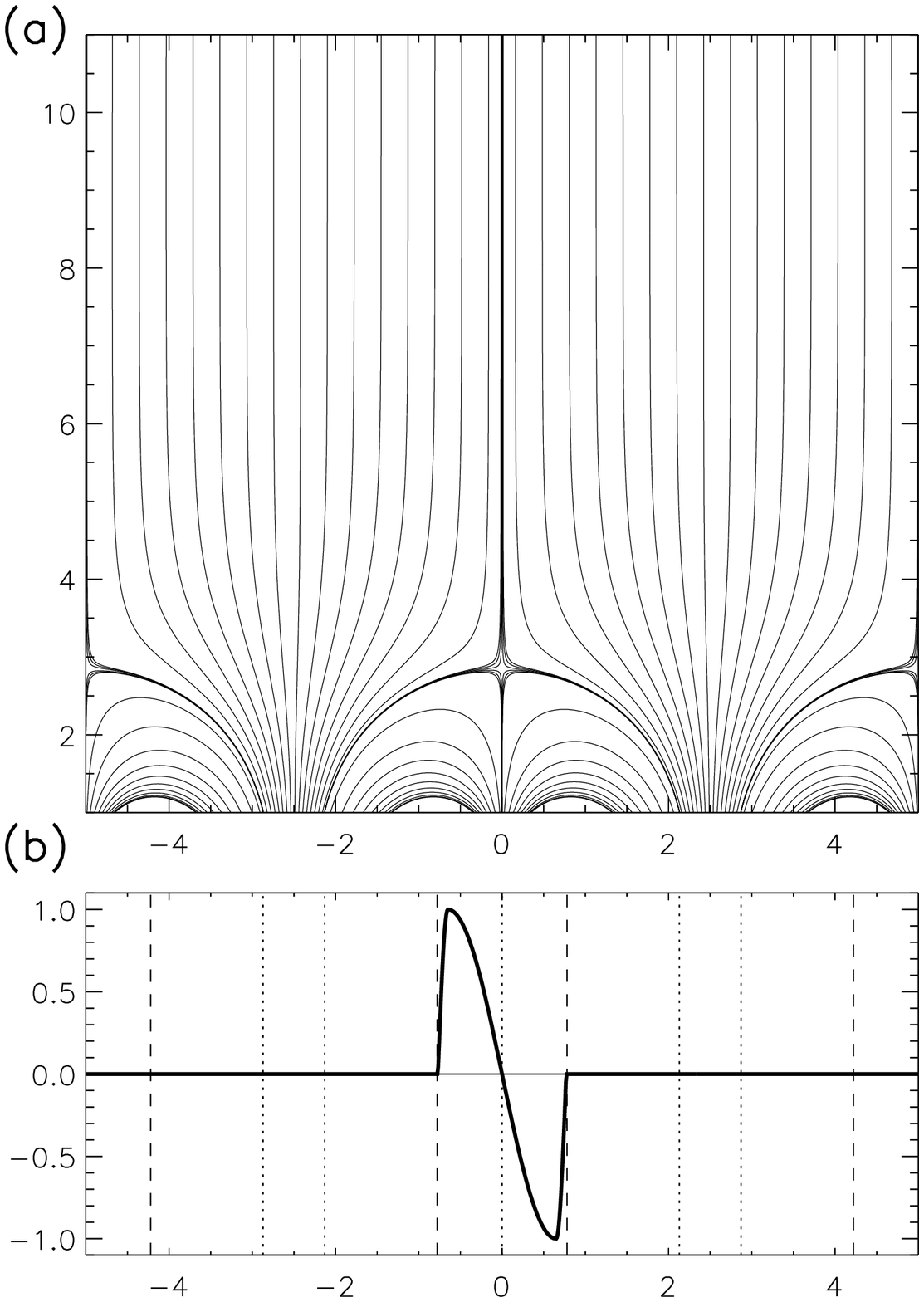}
\caption{Panel (a) shows the initial magnetic field configuration for a pseudostreamer in a background unipolar vertical field. Panel (b) shows the spatial dependence of the shearing flows, the location of the initial separatrix locations (dotted) and polarity inversion lines (dashed). \label{fig1}
}
\end{figure}

The initial magnetic field configuration is constructed via the vector potential for a uniform background field and an infinite series of line dipoles as described in \citet[][]{Edmondson2010a} where we have taken the first 25 terms in the series to ensure a purely vertical field far from the pseudostreamer arcades.
The uniform background field value and line dipole magnitudes are chosen to be \{$B_{y0}$, $M$\} = \{3.0, 30.0\}~G. Figure~\ref{fig1}(a) shows the initial magnetic field topology with representative field lines. 
The null points in each of the initial pseudostreamers occur at $x$=\{0, $\pm 5.0$\} and $y=2.84$. 
While this magnetic configuration is highly idealized, the central arcade's topology is remarkably similar to that obtained in the potential field source surface extrapolation of the pseudostreamer source region of the ``twin filament" eruptions in the 2010 August 1 sequence of eruptive events, analyzed by \citet{Titov2012} and modeled by \citeauthor{Torok2011} (\citeyear{Torok2011}, see their Figure~1(d) and 1(e)).

%

The initial uniform plasma density and pressure are chosen to be $\langle \rho_0 \rangle = 10^{-16}$~gm~cm$^{-3}$ and $\langle P_0 \rangle = 10^{-2}$~dyn~cm$^{-2}$. The initial magnetic energy $E_{\rm M}(0)=9.96\times10^{29}$~erg and total mass $m=10^{14}$~gm in the system yield a globally averaged Alfv\'{e}n speed of $\langle V_A \rangle = ( 2E_{\rm M}(0)/m )^{1/2} = 1411$~km~s$^{-1}$, a characteristic field strength of $\langle B_0 \rangle \sim 5$~G, and a globally averaged plasma beta $\langle \beta _0 \rangle = 8\pi \langle P_0 \rangle /  \langle B_0 \rangle^2 \sim 0.01$.

The system is energized using ideal footpoint shearing flows along the bottom $x$-boundary given by ${\bf V}_{\rm shear} = V_0\left[ F_1(x) G_1(t) + F_2(x) G_2(t)\right] \hat{\bf z}$. Here the subscripts 1, 2 correspond to the shearing profiles associated with each of the two pseudostreamer arcade PILs. The left arcade PIL has 
$F_1(x) = \{  \sin{\left[ \pi (x+0.77)/0.24\right]}$ for $-0.77 \le x < -0.65$; 
                     $\sin{\left[ -\pi x/1.30\right]}$ for $-0.65 \le x < 0.0$; 
                        and 0.0 elsewhere$\}$. 
The right arcade PIL has 
$F_2(x) = \{  \sin{\left[ -\pi x/1.30\right]}$ for $0.0 \le x < 0.65$; 
                     $\sin{\left[ \pi (x-0.77)/0.24\right]}$ for $0.65 \le x < 0.77$; 
                        and 0.0 elsewhere$\}$. 
The spatial dependence of the boundary flows are shown in Figure~\ref{fig1}(b) where the positions of the arcade PILs are shown as vertical dashed lines and the pseudostreamer separatrices are shown as vertical dotted lines. 
The time dependence of the boundary shearing flows are defined as $G_1(t) = \{ 0.5 - 0.5 \cos{\left[ \pi t/100\right]}$ for $ 0.0 \le t < 100$~s; $G_1(t) = 1.0$ for $100 \le t < 1100$~s; and $G_1(t) = 0.5 - 0.5 \cos{\left[ \pi (t-1000)/100\right]}$ for $ 1100 \le t < 1200$~s$\}$. Likewise, $G_2(t)=\{ G_1(t)$ for $ 0.0 \le t < 100$~s; $G_2(t) = 1.0$ for $ 100 \le t < 1250$~s; and $G_2(t) = 0.5 - 0.5 \cos{\left[ \pi (t-1150)/100\right]}$ for $ 1250 \le t < 1350$~s$\}$.
The duration of shearing temporal profiles are shown in Figure~\ref{fig2} ($G_1$, blue; $G_2$, red). 
The maximum magnitude of the shear flows are $V_0 = 75$~km~s$^{-1}$ which is on the order of 5\% of $\langle V_A \rangle$ and roughly 60\% of the characteristic sound speed $( \gamma \langle P_0 \rangle / \langle \rho_0 \rangle)^{1/2} = 129$~km~s$^{-1}$.

Our imposed shearing flows represent a generic process of energy accumulation. Shearing flows in general are both associated with the more complex flux emergence process \citep[e.g.,][]{Strous1996,Magara2003,Manchester2004,Fang2012} and encompass the effects of large scale flux transport processes (differential rotation, meridional flow, granular diffusion) which are a crucial component to the formation and evolution of filament channels \citep[e.g.,][]{vanBallegooijen1998,Yeates2009}. 
High resolution observations of filament/prominence material shows the overall ribbon topology is made up of many fine scale, filamentary strands running parallel to PIL \citep{Martin1998,Lin2008}, precisely the magnetic configuration resulting from a sheared arcade \citep[e.g.,][]{DeVore2000}.
Another advantage to our boundary flows are that they impart shear to both pseudostreamer lobes in the same sense of handedness/chirality. Here, our resulting ``twin filament channel'' sheared arcade fields are both left-handed/dextral, as in the \citet{Panasenco2010} observations and the \citet{Torok2011} simulation. 

%


\section{Simulation Results}
\label{sec_res}

\subsection{Overview}

%

\begin{figure}
\center \includegraphics[width=24pc]{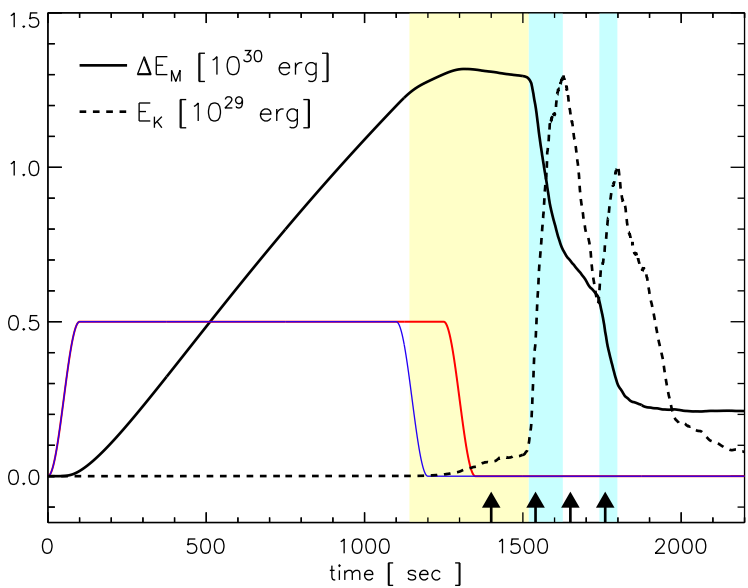}
\caption{The energy evolution of the system: change in magnetic energy $\Delta E_{\rm M}$ (black solid), kinetic energy $E_{\rm K}$ (black dashed). The blue (red) lines indicate the duration of the shearing flows used to energize the pseudostreamer left (right) arcade. The yellow shaded region denotes the breakout reconnection preceding the first CME and the blue shaded regions show the impulsive phase of the eruptive flare reconnection for the first and second CMEs. The arrows denote the times shown in Figure~\ref{fig3}. \label{fig2}
}
\end{figure}

\begin{figure}
\center \includegraphics[width=39pc]{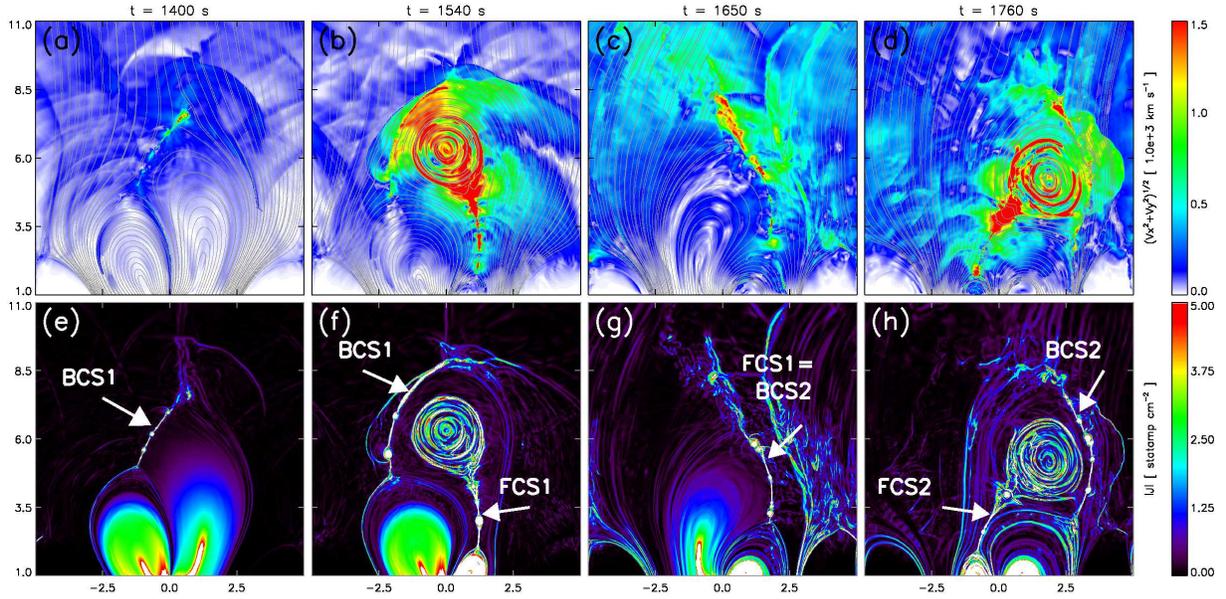}
\caption{Evolution of the sympathetic pseudostreamer eruptions. Top row shows the planar velocity magnitude $V_{xy} = (V_x^2 + V_y^2)^{1/2}$ and the bottom row shows the current density magnitude $J$. The simulation times correspond to: (a) reconnection at the breakout current sheet (BCS1) after the symmetry is broken and the shearing flows turned off, (b) the impulsive phase of reconnection at the first eruptive flare current sheet (FCS1) and CME flux rope formation, (c) the gradual phase of the first CME's flare reconnection acting as the breakout reconnection for the second eruption (FCS1=BCS2), and (d) the impulsive phase of the second CME's flare reconnection (FCS2) and CME flux rope formation. Animations of this figure are available as electronic attachments to the online version.  \label{fig3}
}
\end{figure}

Figure~\ref{fig2} plots the energy evolution of the system.
The black solid line shows the change in magnetic energy $\Delta E_{\rm M} = E_{\rm M}(t) - E_{\rm M}(0)$, where $E_{\rm M}(0)$ is the energy of the initial potential field. The black dashed line indicates the kinetic energy $E_{\rm K}$. 
Figure~\ref{fig3} plots the plane-of-the-sky velocity magnitude $(V_x^2 + V_y^2)^{1/2}$ and current density magnitude $J$ for simulation times $t=\{ 1400, 1540, 1650, 1760 \}$~s. The times of these snapshots are denoted in Figure~\ref{fig2} as the arrows.

During the main shearing (energy accumulation) phase there is slow monotonic rise in the magnetic energy as a significant $B_z$ component is built up in the pseudostreamer arcades and they expand accordingly. We then ramp down the left arcade shearing first, starting at $t=1100$~s, to deliberately break the symmetry of the system to induce the null-point distortion, spine field line separation, and current sheet formation in the \citet{Syrovatskii1981} fashion.

The light yellow shaded region in Figure~\ref{fig2} ($1140 \lesssim t \lesssim 1520$~s) indicates the formation and development of the overlying breakout reconnection current sheet above the right arcade (the source of the first of the sympathetic eruptions), shown in panels (a), (e) of Figures~\ref{fig3}. The first breakout reconnection current sheet is denoted with an arrow as `BCS1' in Figure~\ref{fig3}(e).
In the magnetic breakout model's positive feedback eruption process \citep{Antiochos1999}, the pseudostreamer right arcade's continual expansion becomes more rapid as restraining overlying flux is transferred into both the pseudostreamer left arcade and the background open field. 
The evolution of magnetic reconnection at the breakout current sheet dictates the initial stages of the eruption and the transition from quasi-ideal evolution to a driven, run-away system. The onset of fast reconnection at the breakout current sheet, its associated reconnection jets, and the increased rate of flux transfer denotes the CME onset and inevitable eruption \citep{Karpen2012}.
The breakout-driven eruption phase can also be seen in the gradual kinetic energy increase and is associated with the commonly observed ``slow rise" phase of filament or prominence material and their low lying sheared fields prior to the eruptive flare.

The runaway expansion forms a vertical current sheet deep in the sheared field core. Continued expansion elongates and thins this current sheet and with the onset of eruptive flare reconnection, ushers in the explosive acceleration phase of the CME eruption \citep{Karpen2012}.
The ``impulsive phase" of the first eruption is indicated in Figure~\ref{fig2} as the light blue shaded region ($1520 \lesssim t \lesssim 1625$~s) and is shown as panels (b), (f) in Figure~\ref{fig3}. 
The first CME's eruptive flare current sheet is also indicated as `FCS1' in the current density panel.
The impulsive phase of the eruptive flare corresponds to the rapid release of free magnetic energy and its conversion into kinetic energy, enhanced radiation, particle acceleration, and bulk plasma heating. 
The first CME reaches a maximum kinetic energy of $\Delta E_{\rm K} \sim 1.2\times10^{29}$~erg during the drop in $\Delta E_{\rm M}$ of $\sim$6$\times10^{29}$~erg over $\sim$100~s.
The rapid increase in $E_{\rm K}$ can also be seen as the Alfv\'{e}nic reconnection jet outflows in the velocity magnitude panel corresponding to speeds of $\ge$1500~km~s$^{-1}$ and the acceleration the entire erupting flux rope structure formed {\it during} the flare reconnection process.

The relaxation phase (also called the ``gradual phase" of an eruptive flare) acts to rebuild the arcade in the wake of the CME eruption. In this pseudostreamer configuration, the continued reconnection underneath the first CME simultaneously acts as breakout reconnection for the second CME. 
This is shown in Figure~\ref{fig3}(g) with the label `FCS1=BCS2'.
The restraining flux of the left arcade is processed through the FCS1/BCS2 current sheet, transferring back to the right arcade and into the open field, causing the left arcade to expand in an analogous fashion until the point of generating a second eruptive flare current sheet `FCS2', this time deep in the sheared core of the left arcade (the $1740 \lesssim t \lesssim 1800$~s blue region in Figure~\ref{fig2} and panels (d), (h) in Figure~\ref{fig3}. 
The second flare ``impulsive phase" (CME explosive acceleration phase) has an increase of $\Delta E_{\rm K} \sim 4.5\times10^{28}$~erg during a drop in $\Delta E_{\rm M}$ of  $\sim$3.2$\times10^{29}$~erg. 
While the sympathetic eruption is less energetic overall, the Alfv\'{e}nic flare reconnection jets again provide $\ge$1500~km~s$^{-1}$ outflows and rapid acceleration. 
For both CMEs, the efficiency of the conversion of released magnetic energy into kinetic energy (20\% and 14\%, respectively) are consistent with previous 2.5D and 3D simulation results \citep{Lynch2008,Reeves2010,Karpen2012}.

\subsection{Current Sheet and Reconnection Properties}



Our simulation resolution is sufficient to resolve the dynamics and evolution of fine-scale structure in both the breakout and eruptive flare current sheets. 
The evolution of the current sheet breakup, formation of the magnetic islands and their ejection can be seen in all of the velocity and current density quantities of Figure~\ref{fig3}, in the plasma density and its running difference signatures of Figure~\ref{fig5}, and in their animations included as electronic attachments to the online version: {\tt f3\_velxy.mp4}, {\tt f3\_jmag.mp4}, {\tt f5\_dens.mp4}, and {\tt f5\_rundiff.mp4}. 
Both the breakout and eruptive flare current sheets are sufficiently long ($\ell \gtrsim 2.5\langle L_0 \rangle$) compared to their width ($a \sim 0.02-0.05\langle L_0 \rangle$) for the aspect ratio $\ell/a \sim 50-100$ to exceed the threshold for the resistive tearing mode instability and the elongated current sheet structures break up into a series of alternating X- and O-type null points facilitating magnetic island/plasmoid formation \citep[e.g.,][]{Furth1963,Biskamp1993,Loureiro2007}.

There are a number of features of the simulations of \citet{Edmondson2010a} and \citet{Karpen2012} that we also see in the evolution of out current sheet structures here. 
In the steady-state scenario of \citet{Edmondson2010a} the tearing of the elongated current sheet with magnetic island formation was to facilitate flux and mass transfer required to maintain a quasi-equilibrium. Under those driving conditions their global, current sheet-averaged reconnection rates were faster than Sweet-Parker but slower than Petschek. The successive ejection of plasmoids from the ends of the current sheet split the ends into a larger Y-shape and effectively shortened the sheet length temporarily before reforming. We see similar dynamics here.
In the high resolution, adaptively refined magnetic breakout CME simulation by \citet{Karpen2012}, the current sheets are all driven by the dynamics of the large scale global evolution and eruption scenario. While there is no quasi-steady state at any point during our simulation, for certain phases of the reconnection at both the breakout and post-eruption flare current sheets, the reconnection is not associated with drastic energy conversion and the evolution is relatively ``steady". However, during the runaway expansion and explosive eruption phases the system has to respond, under the constraints of the governing MHD equations and the line-tied lower boundary, to the forces associated with the global disruption and rapid reconfiguration of the large scale magnetic fields and their associated flux transfer. The rapid energy conversion during these phases drive the current sheet evolution and reconnection rates to approach a faster, Petschek-like state. 
The formation of X- and O-type nulls were quantified by \citet{Karpen2012}, showing their significant increase in number during the fast reconnection phases, as well as their height-time evolution along the vertical flare current sheet into the erupting flux rope structure and into the post-eruption flare arcade loops.

\begin{figure}
\center \includegraphics[width=24pc]{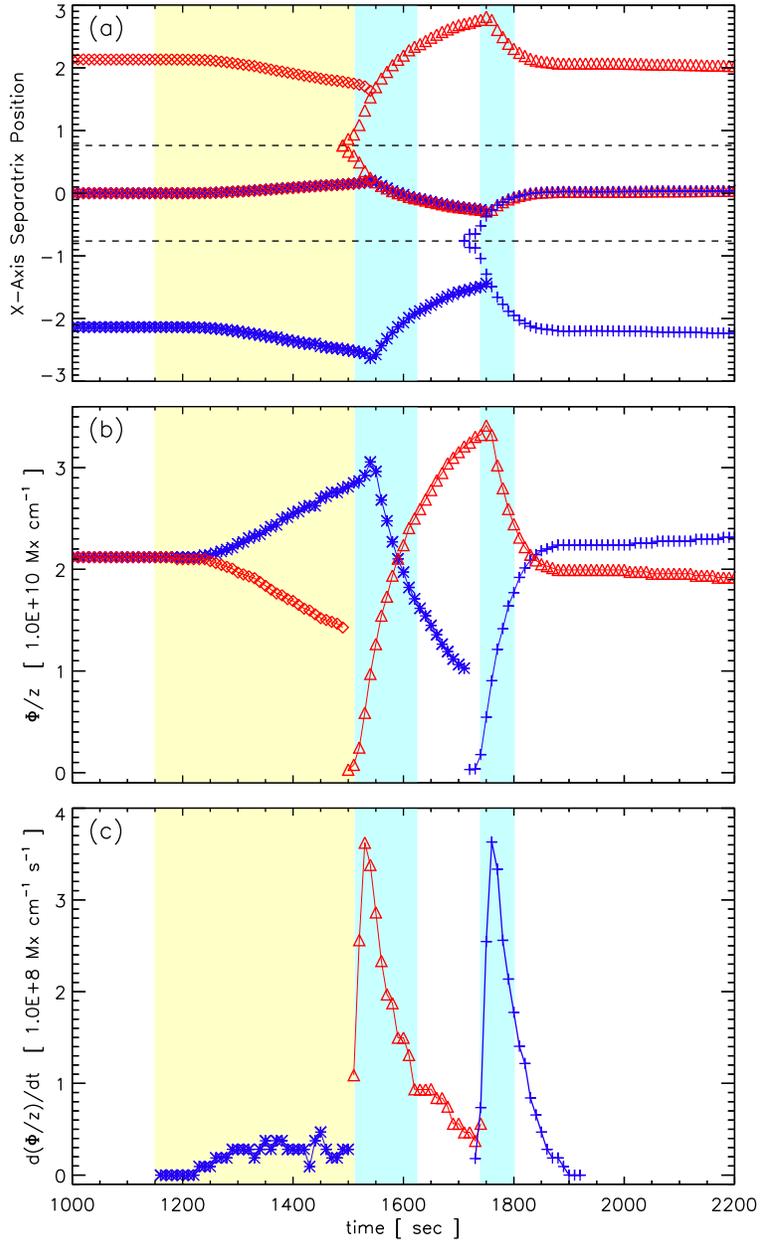}
\caption{Panel (a) plots the $x$-axis position of the pseudostreamer separatricies (red diamonds, blue asterixes) and the post-eruption flare arcades (red triangles, blue crosses). The yellow and light blue shaded regions indicate the breakout reconnection and impulsive flare phases of the sympathetic eruption scenario shown in Figure~\ref{fig2}. Panel (b) plots the flux per unit length contained in each of the flux systems. Panel (c) plots the reconnection rate at the BCS1, FCS1/BCS2, and FCS2 current sheets as the time rate of change of the flux content of the pre-eruption left pseudostreamer arcade and the two post-eruption flare loop systems.    
\label{figrxn}
}
\end{figure}

%
In \citet{Lynch2008} we described the flux transfer process for the magnetic breakout scenario in terms of the evolution of the location of the flux system separatrices on the lower boundary. 
Figure~\ref{figrxn} panel (a) plots the evolution of the position on the $x$-axis of the pseudostreamer flux system boundaries in time. The periods of initial breakout reconnection (the formation and reconnection at BCS1) are highlighted in yellow, while the impulsive phases of fast reconnection in the eruptive flare current sheets FCS1 and FCS2 are shown in blue. The creation of new separatrix surfaces corresponding to the newly created flare arcades show initial rapid separation away from the pseudostreamer PILs and signal the reformation of the arcades in the wake of their respective eruptions. The motion of these separatrix surfaces during the eruptive flare reconnection are associated with the motion of the flare ribbons observed in two-ribbon flares.
Figure~\ref{figrxn} panel (b) plots the temporal evolution of the magnetic flux (per unit length along the $z$ symmetry axis) of the pseudostreamer and flare arcades. The flux per unit length is calculated from $\Phi/z = \int B_y(x,y=1.0)dx$ integrating from the PIL to the separatrix surface locations shown in panel (a). The initial left and right arcades are shown as blue asterixes and red diamonds, respectively, while the flare loop arcades for the first and second CMEs are shown as red triangles and blue crosses. The transfer of flux between the left and right arcades during the first CME's breakout reconnection phase is shown by a nearly linear trend. The post-eruption flare arcade flux content show a very steep rise during their initial formation that slows down and starts to levels off. In the case of the first CME's flare arcade, the interaction with the second CME eruption is clear as flux is rapidly pulled into the second CME's flare current sheet. The time derivative of the flux content is proportional to the reconnection rate, describing how much flux is processed through the current sheet reconnection regions. Panel (c) plots $\partial (\Phi/z)/\partial t$ for the right pseudostreamer arcade during the first breakout reconnection phase (flux through BCS1, blue asterixes) and for the two flare loop arcades (flux through FCS1, red triangles; through FCS2, blue crosses). The various phases of the sympathetic eruption scenario are also clearly distinguishable in the reconnection rate. The breakout reconnection is relatively ``slow" compared to the ``fast" reconnection associated with the onset of the eruptive flares. The time profiles of the flare reconnection rates are qualitatively similar to the hard X-ray profiles associated with the impulsive phase of strong flares \citep[e.g.,][]{Sturrock1980}.


Due to the fine-scale structure in and around the current sheet introduced by the island formation and evolution shown in Figure~\ref{fig3} and the online animations, the localized reconnection rates vary considerably over the length of the current sheet.  
For example, during the impulsive phase of the flare reconnection, the whole upper portion of the current sheet in Figure~\ref{fig3}(f) is experiencing fast reconnection with Alfv\'{e}nic reconnection jet outflows of $\sim$4000~km~s$^{-1}$ and localized inflow speeds on the order of 1000~km~s$^{-1}$. 
Future work will include a quantitative analysis of the reconnection properties including the average and localized inflow and outflow speeds, the variability of the field reversals and plasma conditions along the different current sheets, and the formation, evolution of individual, well-resolved magnetic islands and their role in facilitating mass, energy, and flux transfer through the reconnection region.


\section{Discussion}
\label{sec_disc}

Our 2.5D simulation is highly idealized and so both the structure of the background field and the periodicity affect the CME dynamics considerably. 
While the eruption process itself is well resolved and physical, the subsequent eruption dynamics of the CME flux ropes are subject to the limitations of this simulation. Without a realistic fall-off in height (radius) of the background field, the erupting flux ropes continue their breakout reconnection and are eventually completely dissolved into the uniform vertical field. Physically, this is easily understood as the most efficient redistribution of the imposed highly concentrated shear throughout the simulation volume in an attempt to minimize the total magnetic energy of the system. In a background field with an exponential height/radial dependence faster than or equal to $r^{-2}$, it is more energetically favorable to eject the flux rope out of the system.
%

The periodicity also means that both eruptions exit one side of the simulation domain and wrap-around to the other side. The impact of the remnants of the first eruption on the second eruption's magnetic breakout current sheet development is not important because the pseudostreamer's left arcade expansion and breakout reconnection feedback loop is already well underway (as seen in Figure~\ref{fig3}(c), \ref{fig3}(g)). In the current density movie, {\tt f3\_jmag.mp4}, the first eruption's leading edge is seen to immediately reconnect upon impact with the current sheet facilitating the second CME's breakout reconnection resulting in this flux being transferred out of the way of the sympathetic eruption. 
Despite the inherent limitations of the simulation, there are a number of important features of the results that have implications for our understanding of CME initiation, the role the topology of pseudostreamer source regions play in the eruption process, and predictions of properties that may be observable as pre-eruption signatures from these particular source regions.

%

\subsection{Pre-Eruption Signatures}


The topology of the pseudostreamer means the breakout reconnection transfers flux into the open field directly through interchange reconnection \citep[e.g.,][]{Wang2004,Wang2007,Edmondson2010b,Masson2012} and therefore could potentially give rise to observable pre-eruption signatures in coronagraph observations of pseudostreamers.
In the simulation we see an entire train of plasmoid ejections from the initial breakout reconnection current sheet creating a pre-eruption jet-like outflow up the pseudostreamer stalk as well as blob-like density downflows in the adjacent pseudostreamer arcade. 
The breakout reconnection jet's bursty outflow of material into the open field might have a distinct coronagraph signature (especially in running-difference) as higher variability or higher frequency intensity fluctuations in the pseudostreamer stalks distinguishable from the nominal steady-state solar wind profile.

Observation of downflows in coronagraph, EUV, and X-ray imaging have traditionally been associated with plasma and magnetic field dynamics beneath the eruptive flare current sheet \citep[see][and references therein]{Savage2012} and modeled as such \citep[e.g.,][]{Linton2009,Longcope2009}. While we certainly have those signatures here, we also see them associated with pre-eruption breakout reconnection. Flux transferred from the runaway expansion half of the pseudostreamer to the adjacent arcade is supplying intermittent density enhancements that stream down the magnetic field lines. The appearance of such downflows should precede and coincide with the slow rise phase of the filament undergoing eruption.

\begin{figure}
\center \includegraphics[width=39pc]{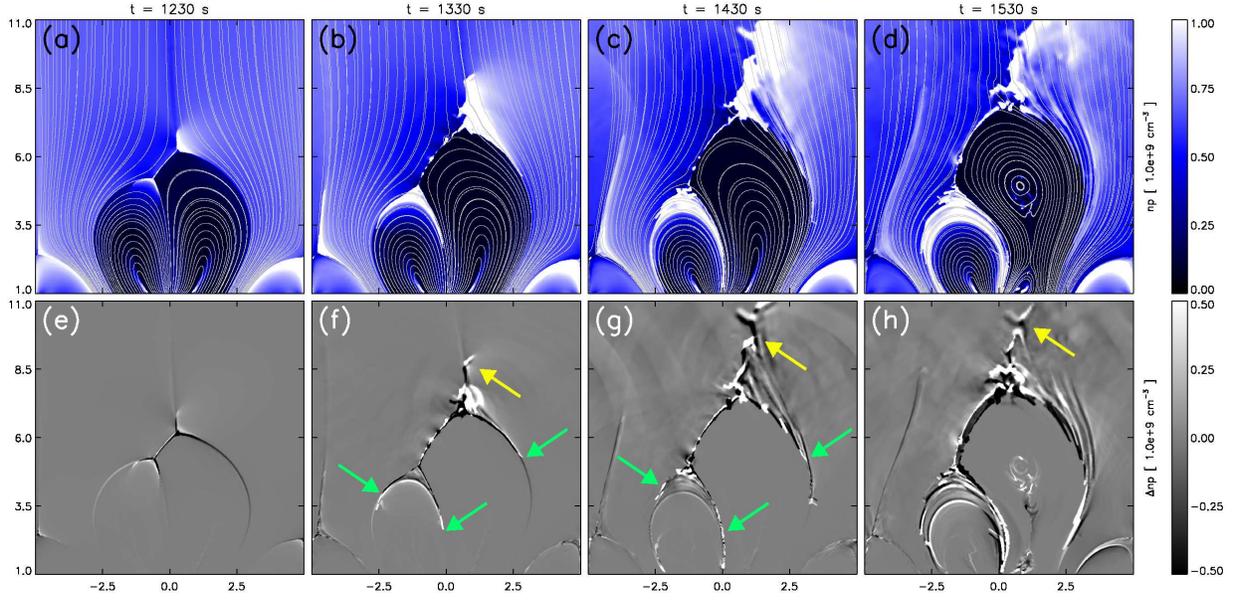}
\caption{Pre-eruption density signatures. Top row shows the plasma number density $n_p$ for (a) $t=1230$~s, (b) $t=1330$~s, (c) $t=1430$~s, and (d) $t=1530$~s. Bottom row panels (e)--(h) show the running difference density signal $\Delta n_p$ for the simulation times above. The yellow (green) arrows point to the upflow (downflow) signatures described in the text. Animations of this figure are available as electronic attachments to the online version. 
\label{fig5}
}
\end{figure}

Figure~\ref{fig5} top row, panels (a)--(d), show the plasma number density $n_p = \rho/m_p$ in 100~s intervals between 1230 and 1530~s immediately preceding the first CME eruption. The bottom row, (e)--(h), show the running difference density signature, $\Delta n_p = n_p(t)-n_p(t-10)$, at each of the above times. 
The yellow arrows point to the overlying breakout reconnection exhaust that forms the interchange reconnection outflows. Because the reconnection becomes bursty due to the current sheet tearing and magnetic island formation, the rising outflow front of density enhancement along the vertical background field has a ragged, fine-scale structure. This interchange reconnection outflow front has a complex running difference signature and each new magnetic island (density blob) absorbed into the open field creates a new alternating positive-negative signal. 
The green arrows point to the running difference downflows in the adjacent pseudostreamer arcade and the newly interchange-reconnected open fields. The downflows are more collimated than the upflows and are seen tracing the (growing) boundaries of the adjacent arcade and adjacent open fieldlines. The running difference signatures are relatively narrow alternating positive-negative fronts in regions of vertical field at the edges of the non-erupting flux system and outline the entire arcade loops as they are compressed under the new loops.
In the wake of CMEs, converging downflows are observed in coronagraphs and EUV and X-ray images at the edge of streamers and near the periphery of active regions \citep{Sheeley2004, Savage2012}, precisely where they are located in our simulation results.  
The online animations of Figure~\ref{fig5} show the entire eruption process in both number density and running difference ({\tt f5\_dens.mp4}, {\tt f5\_rundiff.mp4}).


It will also be interesting to model the bulk plasma heating associated with the breakout reconnection because the plasma material supplied to the adjacent arcade may also have EUV or X-ray emission signatures that precede the initial eruption. In fact, the STEREO/EUVI signatures of the loops above the central arcade in the well-studied 2007 May 19 CME showed exactly this feature: the pre-eruption brightening and arcade growth (piling on of new loops) before the eruption of a filament above the PIL of the western arcade of the multipolar flux system \citep{Li2008}. Additionally, heating during both the breakout and eruptive flare reconnection is important because the coronal density and temperature history of the CME and its surrounding material is imprinted in the spatial variation and complexity of interplanetary CME heavy ionic charge states \citep{Lynch2011,Lepri2012}. Recent analyses by \citet{Reinard2012} suggested that the spatial distribution of in-situ iron charge states in the consecutive 2007 May 21--23 ICMEs were consistent with a sympathetic eruption scenario.


\subsection{CME Deflection During Eruption}


The deflection of CMEs during their evolution has been a topic of recent observational and theoretical study \citep[e.g.,][]{Byrne2010,Lugaz2011,Kay2012,Zuccarello2012}. 
While it has been established, particularly during solar minimum conditions, that CMEs tend to be deflected towards the heliospheric current sheet \citep{Cremades2004}, recent 
observational data by \citet{Gui2011} appears to support the hypothesis of \citet{Shen2011} that CME deflection can be predicted and characterized by the local gradient of the magnetic field energy $-{\bf \nabla} B^2$. 
In the model proposed by \citet{Shen2011} the magnetic energy density in the region immediately surrounding the flux rope CME is evaluated, and if, on average, there exists an imbalance in the magnetic energy density, then one might expect a tendency for deflection towards the region of lower magnetic pressure. 
The breakout current sheet and associated separatrix surrounding each CME flux rope include a very well defined $B^2$ minimum. The Figure~\ref{fig3} movies show the breakout current sheets aligning themselves with the vertical background field so the $-{\bf \nabla} B^2$ maximum becomes parallel to the $x$-axis in both eruptions. The fact that the pseudostreamer topology necessarily includes a null point and flux rope separatrix early in the eruptions could mean pseudostreamer CMEs are more susceptible to deflections of this kind.

\begin{figure}
\center \includegraphics[width=39pc]{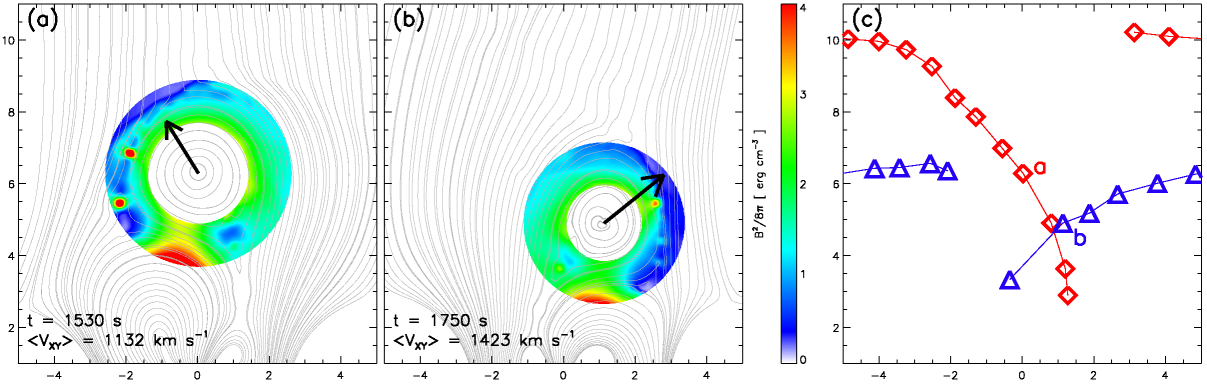}
\caption{Panels (a) and (b) plot the magnetic energy density $B^2/(8\pi)$ surrounding the flux rope CMEs early in their eruptions. The black arrows represent the average planar velocity $\langle V_{xy} \rangle$ over the flux rope cross-section. Panel (c) plots the trajectory of the flux rope centers (first CME, red; second CME, blue) before the continual breakout reconnection dissipates each of the CMEs into the background field. Both CMEs pass through the periodic boundaries and therefore their trajectories appear discontinuous in the plot.
\label{fig6}
}
\end{figure}

Figure~\ref{fig6} panels (a) and (b) plot two snapshots, one from each CME, relatively early in their eruption phases, but after the creation of the flux rope component. The representative fieldlines show the CME location and topology with respect to the background field. The circular ring surrounding each flux rope plots the smoothed magnetic energy density $B^2/(8\pi)$. The black arrows show the direction of the flux rope averaged planar velocity, $\langle V_{xy} \rangle$, in the unshaded circle at the center of the each ring approximating the flux rope's center of motion. Qualitatively, the breakout current sheet half of the magnetic energy density ring in each panel is clearly lower than in the opposite half.  Panel (c) plots the spatial trajectory of each of the CME's flux rope center during their lifetime, i.e. before the flux rope is entirely consumed/reconnected into the background field. The first CME is shown as red diamonds, the second as blue triangles, and the data points corresponding to panels (a) and (b) are labeled accordingly. 
Here, both CMEs pass through the periodic boundaries and therefore their trajectories appear discontinuous in the plot.
The transition from a more vertical to more horizontal trajectory for the first CME and an almost exclusively horizontal trajectory for the second CME appears consistent with the idea that the energy density minimum associated with the breakout current sheet and its continual reconnection provide a path of least resistance for the CME propagation. The magnetic islands in the current sheet show up as localized hot spots of enhanced energy density and so may act to impede the flux rope propagation, as was seen in the simulation by \citet{MacNeice2004}. However, the island formation and ejection from the reconnection region is highly dynamic and both our results and the \citet{Karpen2012} simulations show the islands are both continually moving out of the way of the erupting CME and are much much smaller than the CME flux ropes.

\citet{Panasenco2012} discussed observations of eruptions from pseudostreamer topologies, including the sympathetic eruptions on 1 August 2010 simulated by \citet{Torok2011}. The deflection (i.e., non-radial propagation) of the prominence material in the early phases of eruption were quantified and showed, at least in some cases, the low coronal trajectories are consistent with being ``guided towards weaker field regions, namely null points existing at different heights in the overlying configuration." \citet{Panasenco2012} also discussed the role of asymmetric forces, including those resulting from the large scale coronal hole structure, in the eruption process. \citet{Kay2012} have examined the balance between magnetic tension and pressure-gradient terms and have shown that, while the magnetic pressure gradient is important, other forces can be of the same magnitude and should also be taken into consideration. The \citet{Torok2011} simulation also shows a significant deflection, particularly of their secondary, sympathetic eruption, similar to the motion of the observed prominence material.


\section{Conclusions}
\label{sec_conc}

We have presented the results of a high resolution 2.5-dimensional MHD simulation of sympathetic magnetic breakout CMEs. Because the magnetic topology of a coronal pseudostreamer is favorable to the breakout CME initiation mechanism, we reproduced the sympathetic eruption aspect of the \citet{Torok2011} multiple flux-rope eruption scenario; specifically, the eruptive flare current sheet of the first pseudostreamer CME acts as the pre-eruption breakout reconnection for the second sympathetic eruption.
While the \citet{Torok2011} simulation required a pair of pre-existing flux ropes in quasi-equilibrium underneath the pseudostreamer and an external trigger to disrupt this equilibrium (which took the form of the eruption of an earlier CME), we have shown that with the application of simple ideal footpoint shearing along the pseudostreamer arcade PILs -- a completely generic energization process -- the initiation of consecutive, sympathetic eruptions result as a consequence of the source region topology. 
We have visualized the complex dynamics of the driven current sheet formation and dissipation via magnetic reconnection and discussed the fine-scale structure of the current sheet fragmentation into a series of X- and O-type nulls, recently investigated by \citet{Karpen2012}.
We quantified the flux transfer between the pseudostreamer arcades and in the creation of the post-eruption flare arcades as well as calculated the reconnection rate at the breakout and eruptive flare current sheets, showing that the fast reconnection during the impulsive phase of the eruption corresponds to the rapid conversion of stored magnetic energy into CME kinetic energy and acceleration.
The topology of the coronal pseudostreamer source region may give rise to both observable signatures of the overlying breakout reconnection process that precede the impulsive flare and filament eruption in the form of downflows in the adjacent non-erupting pseudostreamer arcade as well as reconnection-driven outflow along the pseudostreamer arcade stalk. Additionally, the unipolar open field on either side of the pseudostreamer will result in a CME topology early in the eruption that includes a current sheet and magnetic null point (or series of null points) along the field-reversal interface between the flux rope and background coronal field.  Continued breakout reconnection in the low corona when coupled with the natural magnetic energy density minimum of this topology could play a significant role in the observed deflection of high latitude CMEs during the eruption process.

\acknowledgments

The authors thank the reviewer for helpful comments and suggestions that substantially improved the paper. The authors also appreciate valuable discussion with C.~R.~DeVore and S.~K.~Antiochos during the preparation of the manuscript. B.J.L. acknowledges support from AFOSR YIP FA9550-11-1- 0048 and NASA HTP NNX11AJ65G. J.K.E. acknowledges support from NASA LWS NNX10AQ61G and NNX07AB99G.




%

\clearpage



%

\end{document}